\documentstyle[12pt]{article}
\newenvironment{namelist}[1]{%
\begin{list}{}
         {
           \settowidth{\labelwidth}{#1}
           \setlength{\leftmargin}{1.1\labelwidth}
          }
        }{%
\end{list}}   

\newcommand{\bdc}{
\begin{document}}
\newcommand{\edc}{
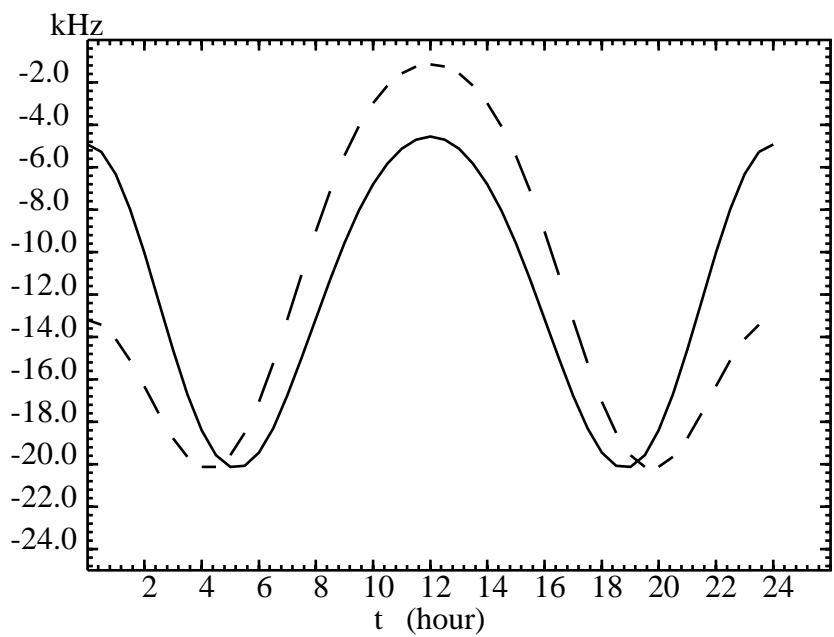
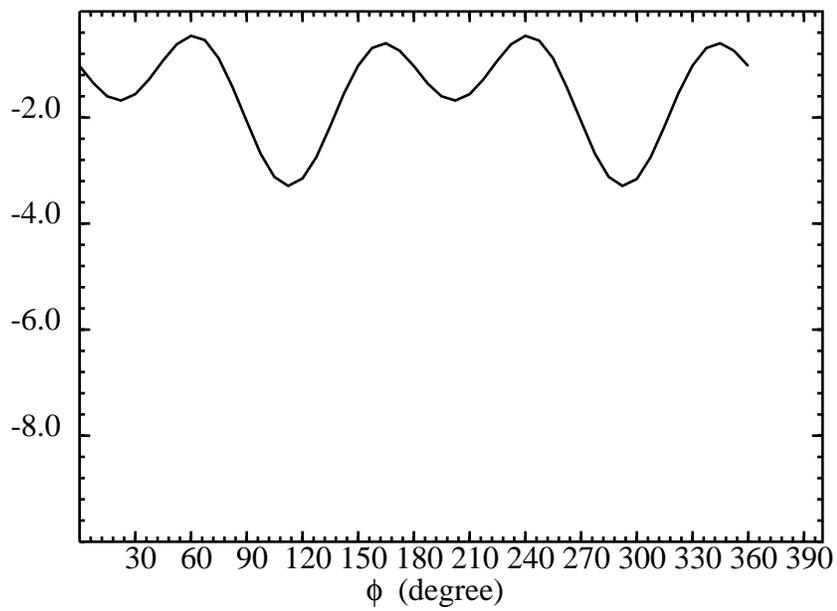
\end{document}}
\newcommand{\bea}{\begin{array}}
\newcommand{\eea}{\end{array}}
\newcommand{\beb}{\begin{thebibliography}}
\newcommand{\eeb}{\end{thebibliography}}
\newcommand{\bec}{\begin{center}}
\newcommand{\eec}{\end{center}}
\newcommand{\bed}{\begin{namelist}}
\newcommand{\eed}{\end{namelist}}
\newcommand{\bee}{\begin{enumerate}}
\newcommand{\eee}{\end{enumerate}}
\newcommand{\bef}{\begin{figure}}
\newcommand{\eef}{\end{figure}}
\newcommand{\beq}{\begin{equation}}
\newcommand{\eeq}{\end{equation}}
\newcommand{\bep}{\begin{picture}}
\newcommand{\eep}{\end{picture}}
\newcommand{\bet}{\begin{tabular}}
\newcommand{\eet}{\end{tabular}}

\newcommand{\bel}{\begin{flushleft}}
\newcommand{\eel}{\end{flushleft}}
\newcommand{\ber}{\begin{flushright}}
\newcommand{\eer}{\end{flushright}}

\newcommand{\rgr}{\raggedright}
\newcommand{\rgl}{\raggedleft}
\newcommand{\ep}{\epsilon}
\newcommand{\vep}{\varepsilon}
\newcommand{\Ep}{\Epsilon}
\newcommand{\ib}{{\it ibid. }}
\newcommand{\af}{\alpha}
\newcommand{\bc}{\bigcirc}
\newcommand{\kp}{\kappa}
\newcommand{\dt}{\delta}
\newcommand{\Dt}{\Delta}
\newcommand{\gm}{\gamma}
\newcommand{\Gm}{\Gamma}
\newcommand{\ld}{\lambda}
\newcommand{\Ld}{\Lambda}
\newcommand{\og}{\omega}
\newcommand{\Og}{\Omega}
\newcommand{\bbt}{\bibitem}
\newcommand{\sgm}{\sigma}
\newcommand{\Sgm}{\Sigma}
\newcommand{\sta}{\theta}
\newcommand{\Sta}{\Theta}
\newcommand{\ups}{\upsilon}
\newcommand{\Ups}{\Upsilon}
\newcommand{\mtc}{\multicolumn}
\newcommand{\lra}{\leftrightarrow}
\newcommand{\ptl}{\partial}
\newcommand{\xrt}{\rightarrow}
\newcommand{\dpst}{\displaystyle}

\newcommand{\ptv}{\stackrel{\leftrightarrow}{\partial}}
\newcommand{\gfv}{\gamma_5}
\newcommand{\gmu}{\gamma^\mu}
\newcommand{\gnu}{\gamma^\nu}
\newcommand{\hf}{\frac{1}{2}}
\newcommand{\qf}{\frac{1}{4}}
\newcommand{\xb}[1]{\overline{#1}}
\newcommand{\xd}[1]{\underline{#1}}
\newcommand{\xcl}[1]{{\cal #1}}
\newcommand{\xp}[1]{{#1}^\prime}
\newcommand{\xpp}[1]{{#1}^{\prime\prime}}
\newcommand{\xs}[1]{#1\!\!\!\!\!\not\;\,}
\newcommand{\Xs}[1]{#1\!\!\!\!\!\!\!\not\;\;}
\newcommand{\Xss}[1]{#1\!\!\!\!\!\!\not\;\;}
\newcommand{\td}[1]{\tilde{#1}}
\newcommand{\dg}[1]{{#1}^\dagger}
\newcommand{\rf}[1]{$^{[#1]}$}
\newcommand{\vt}[1]{\frac{1}{#1}}

\newcommand{\sw}{\sin\theta_W}
\newcommand{\swq}{\sin^2\theta_W}
\newcommand{\cw}{\cos\theta_W}
\newcommand{\cwq}{\cos^2\theta_W}
\newcommand{\cl}{{\cal L}}
\newcommand{\xcm}{{\cal M}}
\newcommand{\abp}{|\vec{p}|}
\newcommand{\xgg}{\gamma\gamma}
\newcommand{\ppw}{pp\rightarrow W\gamma\gamma+n-jet+X}
\newcommand{\ppwd}{pp\rightarrow W(\rightarrow l\nu)
\gamma\gamma+n-jet+X}
\newcommand{\ppwp}{pp\rightarrow W^+\gamma\gamma+n-jet+X}

\bdc

\newcommand{\ucp}{1+u\cos(\phi-\theta)}
\newcommand{\ucm}{1-u\cos(\phi+\theta)}
\newcommand{\ldn}{\lambda_\nu}

\bec
{\large\bf An Implication of Ether Drift}

\vspace{1.5cm}
                  Hong-Yi Zhou \\       
       China Center of Advanced Science and Technology (World Laboratory),\\
      P.O.Box 8730,Beijing 100080,China,\\
      and Institute of Modern Physics and Department of Physics,\\Tsinghua 
      University,Beijing 100084,China \footnote{Mailing address} \\

\vspace{2cm}

                           {\large \bf ABSTRACT}\\
\eec
  
       The experimental results of the two-photon absorption(TPA) and 
M\"{o}ssbauer-rotor(MR) for testing the isotropy of the speed of light are 
explained in an ether drift model with a drift velocity of $\sim 10^{-3}c$.  
Further tests of the ether drift 
assumption are suggested.  
        
\newpage
\parindent=30pt

In Einstein's theory of special relativity (SR), the speed of light $c$  
is a universal constant which is isotropic and independent of the source 
velocity in any inertial frame. The advances of modern technology have 
made the precise experimental tests of this fundamental postulate possible. 
The experiment of Riis et al.\rf{1} measured the frequency shift of 
a two-photon transition in a fast atomic beam while the direction of 
the fast beam  is rotated to the fixed stars. This experiment tests the 
isotropy of the first-order Doppler shift.  The M\"{o}ssbauer-rotor 
experiments of Turner and Hill $^{[2]}$ and of Chanpeney et al.$^{[3][4]}$,
measured the  isotropy of the second-order 
Doppler shift of an emitter mounted on 
the rim of a  rotating disk, as received by an absorber at the center. 
These two kinds of experiments are sensitive to the one-way speed of 
light in contrast to the round-trip experiments as the Michelson-Morley 
one.  The experimental results show that the deviations from the special 
relativity seem to be quite small$^{[5]}$. We note that the TPA experiment 
gave a 12-h period frequency variation which was explained by the authors 
to be a systematic effect.  The 12-h period variation also appeared  
in other experiments$^{[6][7]}$. Whether this variation is a systematic 
effect or new physics should be studied carefully . In this paper, 
we show that the TPA 12-h period variation can be explained 
in an ether drift model while the model give a consistant result of 
the MR experiment.  

The basic assumption of our ether drift model is that  the speed of 
light $c$ measured in the rest frame of the ether is  
isotropic and constant. When an atom moves with velocity $\vec{u}$ 
in the ether, it will contract with a factor  $1/\gm$ ($\dpst\gm=\frac
{1}{\sqrt{1-u^2}}$ with $c\equiv 1$) along the  direction  
of $\vec{u}$ and its energy level $E$ will become $E/\gm$ due to  
the retarded potential$^{[8]}$. The atomic transition frequencies 
will all be lowed down by a factor $1/\gm$ which results in the 
time dilation, nonetheless, this change as well as the length 
contraction is not measurable by a comoving observer. That means our 
model is an ether drift model with Lorentz length contraction and 
Larmor time dilation. The round-trip velocity of light 
measured by a comoving observer is 
precisely isotropic and constant. Therefore, the round-trip experiments 
can not distinguish this ether model from the SR.  We may expect that  
some effects different from the SR exist in the one-way experiments 
as can be seen below. 

We consider now the TPA experiment. Assuming the laboratory moves 
with $\vec{u}$ through the ether, the beam velocity relative to the 
laboratory is $\vec{v}$, 
then we have the following laser frequency $\nu_L$:
\beq\bea{l}  
\dpst 
 \nu_L=\frac{1-u\cos(\phi+\theta)}{1-u\cos(\phi+\theta)-v\cos\theta}
\sqrt{\frac{1-u^2-v^2-2uv\cos\phi}{1-u^2}}\nu_1\;,
\eea\eeq         
\beq\bea{l}  
\dpst
 \nu_L=\frac{1+u\cos(\phi-\theta)}{1+u\cos(\phi-\theta)+v\cos\theta}
\sqrt{\frac{1-u^2-v^2-2uv\cos\phi}{1-u^2}}\nu_2\;,
\eea\eeq         
where $\phi$ is the angle between $\vec{u}$ and $\vec{v}$, $\theta$ 
is the angle between the light ray and the beam axis, $\sin\theta=
u\sin\phi$ .  

It is easy to obtain $v$ and $\nu_L$ from eqs.(1) and (2):
\beq\bea{l} 
\dpst
v=\frac{(1-\ld_\nu)}{\cos\theta}\frac{(\ucp)(\ucm)}{\ucp+\ldn(\ucm)}\;,
\eea\eeq 

\beq\bea{l} 
\dpst
\nu_L=\{\frac{(\ucp)(\ucm)}{(\ucp+v\cos\theta)(\ucm-v\cos\theta)}\\
\times \dpst
\frac{1-u^2-v^2-2uv\cos\phi}{1-u^2}\nu_1\nu_2\}^{\frac{1}{2}} \;,
\eea\eeq 
where $\dpst\ldn\equiv\frac{\nu_1}{\nu_2}(<1)$.  If $u=0$, then we get the 
SR result 
\beq\bea{l} 
\dpst
v_0=\frac{1-\ldn}{1+\ldn}=\frac{\nu_2-\nu_1}{\nu_1+\nu_2}\;,\\
\nu_{L0}=\sqrt{\nu_1\nu_2}\;.
\eea\eeq
We also get the SR result $\nu_L=\sqrt{\nu_1\nu_2}$ for $\phi=0,\;
\pi $ with $\dpst v=\frac{v_0(1-u^2)}{1\pm uv_0}$. Note that this $v$ 
is not directly measured. 

Assuming the beam direction is pointed to the north pole, $\vec{u}$ 
has a declination $\dt$, time $t=0$ when $\vec{u}$ and the beam direction 
have the same right ascensions, we can express $\phi$ as 
\beq 
\dpst
\phi=\cos^{-1}(\sin 56^\circ\cos\dt\cos(\omega t)+\cos 56^\circ\sin\dt)\;,
\eeq
where $56^\circ$ is the lab's latitude, $\omega$ is the Earth's rotation 
angular velocity. We calculated $\dt\nu_L=\nu_L-\nu_{L0}$ as a function of 
$t$ with the parameters $u$ and $\dt$, and found that when $u\sim 
1.2\times 10^{-3}$, $\dt\sim 10^\circ$,$\nu_1=5.045\times 10^{11}$kHz 
and $\nu_2=5.081\times 10^{11}$kHz,    
there exists a 12-h period variation 
with an amplitude of about 7.7 kHz (see Fig. 1). 
They are close to the TPA experiment results of 11.8-h 
period and 7.1 kHz amplitude.

Since the M\"{o}ssbauer-rotor experiment gave a more stringent 
constraint on the frequency variation,  one may worry about that 
the ether model will be excluded by this experiment. We do not 
assume here the rigidity of the rotor because there exist absolute 
length contractions. The received frequency $\nu_r$ is found to be 
\beq\bea{l} 
\dpst
\frac{\nu_r}{\nu_e}=\sqrt{\frac{1-u^2-v^2+2uv\sin(\phi-\dt_2)}{1-u^2}}
\frac{1+u\cos(\phi+\dt_1)}{1+u\cos(\phi+\dt_1)+v\sin(\dt_1+\dt_2)}\;,\\
\sin\dt_1=u\sin\phi\;,\;\sin\dt_2=u^2\sin\phi\cos\phi\;,\\
\eea\eeq
where $\phi$ is the angle between the radial direction of the emitter 
and $\vec{u}$, $\dt_1$ is the 
angle  between the radial direction and the light ray, $\dt_2$ is the 
deviation of the angle between the radial direction and $\vec{v}$ from 
$90^\circ$ ($\dt_2\neq 0$ due to the contraction along 
the $\vec{u}$ direction).  

For $u=1.2\times 10^{-3}$, $v=1.2\times 10^{-6}$, we found
\beq 
\dpst
\frac{\nu_r}{\nu_e}=1-\hf v^2+O(10^{-18}). 
\eeq

The variation of $\dpst\frac{\nu_r}{\nu_e}$ is quite below the MR 
experiment constraint ($2\times 10^{-16}$). We plotted $\dpst(\frac{\nu_r}
{\nu_e}-1+\hf v^2)\times 10^{18}$ against $\phi$ in Fig. 2.    

It is easy to understand our results if one knows that the differences
between the model used here and the SR are of order ${(\vec{u}\cdot\vec{v})}
^2$. For the confirmation of the ether drift model,  further experiments 
must be done. For example, one can do the TPA experiment at different time 
of the year to see if there exists a direction correlation between 
the beam and the maxima of the 12-h period. One can also do the same 
experiment by changing the beam direction with $180^\circ$ which is 
equivalent to the case of $\dt=-10^\circ$. In this case, the 24-h 
amplitude will be observable. Finally, it should be noted that the  
direction of $\vec{u}$ may not be coincide with the symmetry axis 
of the $3$ K microwave background radiation, because local gravitation 
forces may influence the ether drift.

\vspace{5cm}
\bec ACKNOWLEDGMENT \eec

I am grateful to the CCAST for use of their computation facilities. 

\newpage

\beb{7}
\bbt{1} E.Riis, L.-U.A.Anderson, N.Bjerre, O.Poulson, S.A.Lee, and J.L. Hall, 
Phys. Rev. Lett. 60 (1988) 81; 62 (1989) 842. 

\bbt{2} K.C.Turner and H.A.Hill, Phys. Rev. 134 (1964) B252 .

\bbt{3} D.C.Champeney, G.R.Isaak, and A.M.Khan, Phys. Lett 7 (1963) 241.

\bbt{4} G.R.Isaak, Phys. Bull. 21 (1970) 255.

\bbt{5} C.M.Will, Phys. Rev. D45 (1992) 403.

\bbt{6} J.D.Prestage, J.J.Bollinger, W.M.Itano, and D.J.Wineland, 
Phys. Rev. Lett. 54 (1985) 2387.

\bbt{7} S.K.Lamoreaux, J.P.Jocobs, B.R.Heckel, F.J.Raab, and E.N.Fortson,
Phys. Rev. Lett. 57 (1986) 3125.

\bbt{8} S.J.Prokhovnik, $<<$Light in Einstein's Universe $>>$ (Reidel, 
Dordrecht, 1985). 

\eeb

\newpage
\parindent=0pt
{\Large FIGURE CAPTIONS:}
\bef[h]
\caption{ Predicted frequency variations of the TPA experiment 
with $u=1.2\times 10^{-3},\;\;\dt=10^\circ$(solid line) 
and $\dt=-10^\circ$(dashed line ).} 

\caption{ Predicted frequency variation of the MR experiment with 
$u=1.2\times10^{-3},\;\;v=1.2\times 10^{-6}$ .} 

\eef 
        
\edc